\documentclass[11pt]{article}

\usepackage{url}
\usepackage[breaklinks,bookmarks]{hyperref}

\usepackage{fullpage}
\usepackage{latexsym}
\usepackage{amssymb}
\usepackage{amsfonts}

\newcommand{\Z}{{\mathchoice
{\hbox{$\sf\textstyle Z\kern-0.4em Z$}} {\hbox{$\sf\textstyle
Z\kern-0.4em Z$}} {\hbox{$\sf\scriptstyle Z\kern-0.3em Z$}}
{\hbox{$\sf\scriptscriptstyle Z\kern-0.2em Z$}}}}
\newcommand{\Com}{{\mathchoice {\setbox0=\hbox{$\displaystyle\rm
        C$}\hbox{\hbox to0pt{\kern0.4\wd0\vrule height0.9\ht0\hss}\box0}}
{\setbox0=\hbox{$\textstyle\rm C$}\hbox{\hbox
to0pt{\kern0.4\wd0\vrule height0.9\ht0\hss}\box0}}
{\setbox0=\hbox{$\scriptstyle\rm C$}\hbox{\hbox
to0pt{\kern0.4\wd0\vrule height0.9\ht0\hss}\box0}}
{\setbox0=\hbox{$\scriptscriptstyle\rm C$}\hbox{\hbox
to0pt{\kern0.4\wd0\vrule height0.9\ht0\hss}\box0}}}}

\newcommand{\dom}{X\times Y }

\newcommand{\kdom}{X^k\times Y^k }

\newcommand{\lrdom}{X^l\times Y^r }

\newcommand{\fdom}{X\times Y\to\{0,1\} }

\newtheorem{definition}{Definition}
\newtheorem{theorem}{Theorem}
\newtheorem{lemma}{Lemma}
\newtheorem{claim}[lemma]{Claim}
\newtheorem{fact}[lemma]{Fact}
\newtheorem{proposition}[lemma]{Proposition}
\newtheorem{corollary}{Corollary}

\newenvironment{proof}[1][Proof.]{
    \par
    \noindent \textbf{#1}
}{
    \nobreak\leavevmode
    \hfill $\Box$\par\bigskip
}

\date{}
\begin{document}

\title{Quantum and Classical Communication-Space Tradeoffs from Rectangle Bounds}

\author{Hartmut Klauck
\thanks {Supported by DFG grant KL 1470/1. Work partially done
at Department of Computer Science, University of Calgary, supported by Canada's NSERC and
MITACS.}\\
Institut f\"ur Informatik\\
       Goethe-Universit\"at Frankfurt\\
       60054 Frankfurt am Main, Germany\\
       {\tt klauck@thi.informatik.uni-frankfurt.de}
}

\maketitle

\begin{abstract}
We derive lower bounds for tradeoffs between the communication $C$ and space $S$ for
communicating circuits. The first such bound applies to quantum circuits. If for any
problem $f:\dom\to Z$ the multicolor discrepancy of the communication matrix of $f$ is
$1/2^d$, then any bounded error quantum protocol with space $S$, in which Alice receives
some $l$ inputs, Bob $r$ inputs, and they compute $f(x_i,y_j)$ for the $l\cdot r$ pairs
of inputs $(x_i,y_j)$ needs communication $C=\Omega(lrd \log |Z|/S)$. In particular,
$n\times n$-matrix multiplication over a finite field $F$ requires $C=\Theta(n^3\log^2
|F|/S)$, matrix-vector multiplication $C=\Theta(n^2\log^2 |F|/S)$. We then turn to
randomized bounded error protocols, and derive the bounds $C=\Omega(n^3/S^2)$ for Boolean
matrix multiplication and $C=\Omega(n^2/S^2)$ for Boolean matrix-vector multiplication,
utilizing a new direct product result for the one-sided rectangle lower bound on
randomized communication complexity. These results imply a separation between quantum and
randomized protocols when compared to quantum bounds in \cite{ksw:dirprod} and partially
answer a question by Beame et al.~\cite{beame:commtrade}.
\end{abstract}

\section{Introduction}
\label{sec-intro}
\subsection{Quantum Tradeoffs}
Computational tradeoff results show how spending of one resource must be increased when
availability of another resource is limited in solving computational problems. Results of
this type have first been established by Cobham~\cite{cobham:tradeoff}, and have been
found to describe nicely the joint behavior of computational resources in many cases.
Among the most important such results are time-space tradeoffs, due to the prominence of
these two resources. It can be shown that e.g.~(classically) sorting $n$ numbers requires
that the product of time and space is $\Omega(n^2)$~\cite{beame:tradeoff}, and time
$O(n^2/S)$ can also be achieved in a reasonable model of computation for all $\log n\le
S\le n/\log n$ \cite{pagter:tradeoff}.

The importance of such results lies in the fact that they capture the joint behavior of
important resources for many interesting problems as well as in the possibility to prove
superlinear lower bounds for tradeoffs, while superlinear lower bounds for single
computational resources can usually not be obtained with current techniques.

Quantum computing is an active research area offering interesting possibilities to obtain
improved solutions to information processing tasks by employing computing devices based
on quantum physics, see e.g.~\cite{nc:qc} for a nice introduction into the field. Since
the number of known quantum algorithms is rather small, it is interesting to see which
problems might be candidates for quantum speedups. Naturally we may also consider
tradeoffs between resources in the quantum case. It is known that e.g.~quantum time-space
tradeoffs for sorting are quite different from the classical tradeoffs, namely
$T^2S=\widetilde{\Theta}(n^3)$ \cite{ksw:dirprod} (for an earlier result see
\cite{aaronson:advice}). This shorthand notation is meant as follows: the lower bound
says that for all $S$ any algorithm with space $S$ needs time
$\widetilde{\Omega}(n^{3/2}/\sqrt{S})$, while the upper bound says that (in this case for
all $\log^3 n\le S\le n$) there is a space $S$ algorithm with time
$\widetilde{O}(n^{3/2}/\sqrt{S})$.

Communication-space tradeoffs can be viewed as a generalization of time-space tradeoffs.
Study of these has been initiated in a restricted model by Lam et
al.~\cite{lam:commtrade}, and several tight results in a general model have been given by
Beame et al.~\cite{beame:commtrade}. In the model they consider two players only
restricted by limited workspace communicate to compute a function together. Note that
whereas communication-space tradeoffs always imply time-space tradeoffs, the converse is
not true: e.g.~if players Alice and Bob receive a list of $n$ numbers with $O(\log n)$
bits each, then computing the sorted list of these can be done deterministically with
communication $O(n\log n)$ and space $O(\log n)$.

Most of the results in this paper are related to the complexity of matrix multiplication.
The foremost question of this kind is of course whether quantum algorithms can break the
current barrier of $O(n^{2.376})$ for the time-complexity of matrix multiplication
\cite{coppersmith:matrix} (it has recently been shown that {\it checking} matrix
multiplication is actually easier in the quantum case than in the classical case, and can
be done in time $O(n^{5/3})$ \cite{BS:qverif}). In this paper we investigate the
communication-space tradeoff complexity of matrix multiplication and matrix-vector
multiplication. Communication-space tradeoffs in the quantum setting have recently been
established \cite{ksw:dirprod} for {\it Boolean} matrix-vector product and matrix
multiplication. In the former problem there are an $n\times n$ matrix $A$ and a vector
$b$ of dimension $n$ (given to Alice resp.~to Bob), and the goal is to compute the vector
$c=Ab$, where $c_i=\vee_{j=1}^n \left(A[i,j]\wedge b_j\right)$. In the latter problem of
Boolean matrix multiplication two matrices have to be multiplied with the same type of
Boolean product. The paper \cite{ksw:dirprod} gives tight lower and upper bounds for
these problems, namely $C^2S=\widetilde{\Theta}(n^5)$ for Boolean matrix multiplication
and $C^2S=\widetilde{\Theta}(n^3)$ for Boolean matrix-vector multiplication.

Here we first study these problems in the case when the matrix product is not defined by
for the Boolean operations $\wedge$ and $\vee$ (which form a semiring with $\{0,1\}$),
but over finite fields, and again for quantum circuits. Later we go back to the Boolean
product and study the classical complexities of these problems, in order to get a
quantum/classical separation for the Boolean case. All these results are collected in the
following table.

\vspace{.5cm}
\begin{tabular}{|l|l|l||l|l|} \hline
& Fields $F$ & Fields $F$& Boolean & Boolean\\
& Matrix Mult.& Matrix-Vector &Matrix Mult.& Matrix-Vector \\ \hline \hline
Quantum &&&&\\
upper & $O(n^3 \log^2|F|/S)$ & $O(n^2\log^2|F|/S)$ & $\widetilde{O}(n^{5/2}/\sqrt{S})$ & $\widetilde{O}(n^{3/2}/\sqrt{S})$ \\
bound& obvious & obvious & \cite{ksw:dirprod} & \cite{ksw:dirprod}\\ \hline
Quantum&&&&\\
lower & $\Omega(n^3 \log^2|F|/S)$ & $\Omega(n^2\log^2|F|/S)$ & $\Omega(n^{5/2}/\sqrt{S})$
& $\Omega(n^{3/2}/\sqrt{S})$ \\
bound& this paper& this paper
&\cite{ksw:dirprod}&\cite{ksw:dirprod}\\ \hline
Deterministic &&&&\\
upper & $O(n^3 \log^2|F|/S)$ & $O(n^2\log^2|F|/S)$ & $O(n^{3}/ S)$ & $O(n^{2}/S)$ \\
bound& obvious & obvious & obvious & obvious\\
 \hline
Randomized&&&&\\
lower& $\Omega(n^3 \log^2|F|/S)$ & $\Omega(n^2\log^2|F|/S)$ & $\Omega (n^{3}/ S^2)$ & $\Omega(n^{2}/S^2)$ \\
bound&\cite{beame:commtrade} &\cite{beame:commtrade}&this paper& this paper\\
 \hline\end{tabular}
\vspace{.5cm}

Note that in the above table all upper bounds hold for $\log n\le S\le n$, and that the
results from \cite{beame:commtrade} are actually shown in a slightly different model
(branching programs that communicate field elements at unit cost) and hence stated with a
factor of $\log |F|$ less there.

\subsection{Direct Product Results}
As in \cite{ksw:dirprod} we use direct product type results to obtain quantum
commu\-nication-space tradeoff lower bounds for functions with many outputs. In this
approach (as in previous proofs concerning such tradeoffs) a space bounded circuit
computing a function is decomposed into slices containing a certain amount of
communication. Such a circuit slice starts with a (possibly complicated) initial state
computed by the gates in previous slices, but this state can be replaced by the totally
mixed state at the cost of reducing the success probability by a factor of $1/2^S$, where
$S$ is the space bound. If we manage to show that a circuit with the given resources (but
with no initial information) can compute $k$ output bits of the function only with
success probability exponentially small in $k$, then $k=O(S)$, and we can prove a
tradeoff result by concluding that the number of circuit slices times $O(S)$ must be
larger than the number of output bits.

A direct product result says that when solving $k$ instances of a problem simultaneously
the success probability will go down exponentially in $k$. There are two different types
of direct product results. In a {\it strong} direct product result we try to solve $k$
instances with $k$ times the resources that allow us to solve the problem on one instance
with probability $2/3$. In a {\it weak} direct product theorem we have only the same
amount of resources as for one instance.

Our approach is to show direct product type results for lower bound techniques that work
for quantum resp.~randomized communication complexity of functions $f$. We focus on lower
bound methods defined in terms of the properties of rectangles in the communication
matrix of $f$. There are several techniques available now for proving lower bounds on the
quantum communication complexity (see \cite{razborov:qdisj,klauck:qclb}). The earliest
such technique was the discrepancy bound first applied to quantum communication by Kremer
\cite{kremer:thesis}. This bound is also related to the majority nondeterministic
communication complexity \cite{klauck:qclb}.

\begin{definition} Let $\nu$ be a distribution on
 $\dom$ and $f$ be any function $f:\fdom$.
Then let $disc_\nu(f)=\max_R|\nu(R\cap f^{-1}(0))-\nu(R\cap f^{-1}(1))|,$ where $R$ runs
over all rectangles in the communication matrix of $f$ (see Section~\ref{sec:rect}).
\end{definition}

In the rest of the paper $\mu$ will always denote the uniform distribution on some
domain. $disc(f)$ will be a shorthand for $disc_\mu(f)$. We will also refer to the term
maximized above as the discrepancy of a particular rectangle. Since we are dealing with
multiple output problems, also a notion of {\it multicolor} discrepancy we are going to
define later will be useful. $-\log(disc(f))$ gives a lower bound on the quantum
communication complexity \cite{kremer:thesis}.

As Shaltiel \cite{shaltiel:sdpt} has pointed out, in many cases strong direct product
theorems do not hold. He however gives a strong direct product theorem for the
discrepancy bound, or rather a XOR-lemma: he shows that
\[disc(\oplus_{i=1,\ldots,k} f(x_i))\le disc(f(x))^{\Omega(k)}.\]
Previously Parnafes et al.~\cite{prw:productgcd} showed a general direct product theorem
for classical communication complexity, but in their result the success probability is
only shown to go down exponentially in $k/c$, where $c$ is the communication complexity
of the problem on one instance, so this result cannot be used for deriving good tradeoff
bounds. Klauck et al.~\cite{ksw:dirprod} have recently given a strong direct product
theorem for computing $k$ instances of the Disjointness problem in quantum communication
complexity.

Instead of the usual direct product formulation ($k$ independent instances of a problem
have to be solved) we first focus on the following setup (a generalized form of matrix
multiplication): Alice receives $l$ inputs, Bob receives $r$ inputs, and they want to
compute $f(x_i,y_j)$ for all $lr$ pairs of inputs for some function $f$. We denote this
problem by $f_{l,r}$. We will show that when the communication in a quantum protocol is
smaller than the discrepancy bound (for one instance) then the success probability of
computing some $k$ of the outputs of $f_{l,r}$ goes down exponentially in $k$ (for all
$k$ smaller than the discrepancy bound), and refer to such a result as a {\it bipartite
product result}.  This differs from Shaltiel's direct product result for discrepancy
\cite{shaltiel:sdpt} in three ways: first, it only holds when the communication is
smaller than the discrepancy bound for one instance (like a weak direct product result),
secondly, it deals with correlated input instances (in the described bipartite way).
Furthermore it is not  about discrepancy of the XOR of the outputs for $k$ instances, but
rather about the multicolor discrepancy.

\subsection{Our Results}

The first lower bound result of this paper is the following:

\begin{theorem}\label{the:cst}
Let $f:\fdom$ with $disc(f)\le 1/2^d$. Then any quantum protocol
using space $S$ that computes $f_{l,r}$ needs communication
$\Omega(dlr/S)$.
\end{theorem}

A completely analogous statement can be made for functions
$f:\dom\to Z$ for some set $Z$ of size larger than two and
multicolor discrepancy, where the lower bound is larger by a
factor of $\log |Z|$.

The inner product function over a field $F$ is $IP^F(x,y)=\sum_{i=1}^n x_i\cdot y_i$ with
operations over $F$. $IP^{GF(2)}$ has been considered frequently in communication
complexity theory. It is known that its quantum communication complexity is $\Theta(n)$
(the lower bound can be proved using discrepancy \cite{kremer:thesis}). Note that
$IP^F_{n,n}$ corresponds to the multiplication of two $n\times n$ matrices over $F$,
while $IP^F_{n,1}$ is the matrix-vector product. It is well known that
$disc(IP^{GF(2)})\le 2^{-n/2}$ (see \cite{kushilevitz&nisan:cc}). A generalization of
this result given by Mansour et al.~\cite{mnt:hash} implies similar bounds on the
multicolor discrepancy of inner products over larger fields. Together with a trivial
deterministic algorithm in the model of communicating circuits we get the following
corollary.

\begin{corollary}\label{cor:mmfield}
Assume $\log n\le S\le n\log |F|$.

$IP^F_{n,n}$ can be computed by a deterministic protocol with space $S$ and communication
$O(n^3\log^2 (|F|)/S)$, and any bounded error quantum protocol with space $S$ needs
communication $\Omega(n^3\log^2 (|F|)/S)$ for this problem.

$IP^F_{n,1}$ can be computed by a deterministic protocol with space $S$ and communication
$O(n^2\log^2 (|F|)/S)$, and any bounded error quantum protocol with space $S$ needs
communication $\Omega(n^2\log^2 (|F|)/S)$ for this problem.
\end{corollary}

Using a lemma from \cite{mnt:hash} (also employed in
\cite{beame:commtrade}) we are also able to give a lower bound for
pairwise universal hash functions.

\begin{definition}
A pairwise universal family $Y$ of hash functions from a set $X$
to a set $Z$ has the following properties when $h\in Y$ is chosen
uniformly at random:
\begin{enumerate}
\item For any $x\in X$: $h(X)$ is uniformly distributed in $Z$.
\item For any $x,x'\in X$ with $x\neq x'$, and any $z,z'\in Z$,
the events $h(x)=z$ and $h(x')=z'$ are independent.
\end{enumerate}

In the problem of evaluating a hash function by a protocol Alice
gets $x\in X$, Bob gets a function $h\in Y$, and they compute
$h(x)$.
\end{definition}

\begin{corollary}\label{cor:hash}
Any bounded error quantum protocol that evaluates a pairwise universal family of hash
functions using space $S$ needs communication at least
$\Omega(\min\{\log(|X|)\cdot\log(|Z|)/S\, , \,\log^2(|Z|)/S\}).$
\end{corollary}

Beame et al.~\cite{beame:commtrade} have established the first term in the above
expression as a lower bound for randomized communicating circuits. Hence our quantum
lower bound is weaker for hash functions that map to a small domain.

There are many examples of pairwise universal hash function, see \cite{mnt:hash}. Let us
just mention the function $f:GF(r)\times GF(r)^2\to GF(r)$ defined by $f(x,(a,b))=a\cdot
x+b$. If $n=\lceil\log r\rceil$ then this function has a quantum communication tradeoff
$CS=\Omega(n^2)$. Also there are universal hash functions that can be reduced to
matrix-multiplication and matrix-vector multiplication over finite fields, and we could
have deduced the result about matrix-vector multiplication in Corollary~\ref{cor:mmfield}
from the above result. The result about matrix multiplication would not follow, since the
standard reduction from convolution (see \cite{mnt:hash}, matrix multiplication itself is
not a hash function) has the problem that for convolution the $\log^2|Z|$ term is much
smaller than the $\log |X|\cdot \log |Z|$ term, and we would not get a good lower bound.
Also not every function $f_{l,r}$, where $f$ has small discrepancy, is a universal hash
function.

We then turn to classical communication-space tradeoffs for Boolean matrix and Boolean
matrix-vector multiplication. We show a weak direct product theorem for the one-sided
rectangle bound on randomized communication complexity, which allows us to deduce a weak
direct product theorem for the classical complexity of the Disjointness problem. Using
this we can show a communication-space tradeoff lower bound for Boolean matrix
multiplication, a problem posed by Beame et al.~\cite{beame:commtrade}.

In the Disjointness problem Alice has an $n$-bit input $x$ and Bob has an $n$-bit input
$y$. These $x$ and $y$ represent sets, and $DISJ(x,y)=1$ iff those sets are disjoint.
Note that $DISJ$ is $NOR(x\wedge y)$, where $x\wedge y$ is the $n$-bit string obtained by
bitwise AND-ing $x$ and $y$. The communication complexity of $DISJ$ has been well
studied: it takes $\Theta(n)$ communication in the classical (randomized)
world~\cite{ks:disj,razborov:disj} and $\Theta(\sqrt{n})$ in the quantum world
\cite{BuhrmanCleveWigderson98,hoyer&wolf:disjeq,aaronson&ambainis:search,razborov:qdisj}.
A strong direct product theorem for the {\it quantum} complexity of Disjointness has been
established in \cite{ksw:dirprod}, but the randomized case was left open. $DISJ_{n,n}$ is
(the bitwise negation of) the Boolean matrix product.

\begin{theorem}\label{the:disjdpt}
There are constants $\epsilon,\gamma>0$ such that when Alice and
Bob have $k\le \epsilon n$ instances of the Disjointness problem
on $n$ bits each, and they perform a classical protocol with
communication $\epsilon n$, then the success probability of
computing all these instances simultaneously correct is at most
$2^{-\gamma k}$.
\end{theorem}

An application of this gives a classical communication-space
tradeoff.

\begin{theorem}\label{the:disjtrade}
For the problem $DISJ_{n,n}$ (Boolean matrix multiplication) every randomized space $S$
protocol with bounded error needs communication $\Omega(n^3/S^2)$.

For the problem $DISJ_{n,1}$ (Boolean matrix-vector multiplication) every randomized
space $S$ protocol with bounded error needs communication $\Omega(n^2/S^2)$.
\end{theorem}

The proof is in Appendix~\ref{app:D}. Obvious upper bounds are $O(n^3/S)$
resp.~$O(n^2/S)$ for all $\log n\le S\le n$. No lower bound was known prior to the recent
quantum bounds in \cite{ksw:dirprod}. Note that the known quantum bounds for these
problems are tight as mentioned above. For small $S$ we still get near-optimal separation
results, e.g.~for polylogarithmic space quantum protocols for Boolean matrix
multiplication need communication $\widetilde{\Theta}(n^{2.5})$, classical protocols
$\widetilde{\Theta}(n^3)$. The reason we are able to analyze the quantum situation more
satisfactorily is the connection between quantum protocols and polynomials exhibited by
Razborov \cite{razborov:qdisj}, allowing algebraic instead of combinatorial arguments.

\section{Definitions and Preliminaries}\label{sec-def}

\subsection{Communicating Quantum Circuits}
In the model of quantum communication complexity, two players Alice and Bob compute a
function $f$ on distributed inputs $x$ and $y$. The complexity measure of interest in
this setting is the amount of communication. The players follow some predefined protocol
that consists of local unitary operations, and the exchange of qubits. The communication
cost of a protocol is the maximal number of qubits exchanged for any input. In the
standard model of communication complexity Alice and Bob are computationally unbounded
entities, but we are also interested in what happens if they have bounded memory, i.e.,
they work with a bounded number of qubits. To this end we model Alice and Bob as
communicating quantum circuits, following Yao~\cite{yao:qcircuit}.

A pair of communicating quantum circuits is actually a single
quantum circuit partitioned into two parts. The allowed operations
are local unitary operations and access to the inputs that are
given by oracles. Alice's part of the circuit may use oracle gates
to read single bits from her input, and Bob's part of the circuit
may do so for his input. The communication $C$ between the two
parties is simply the number of wires carrying qubits that cross
between the two parts of the circuit. A pair of communicating
quantum circuits uses space $S$, if the whole circuit works on $S$
qubits.

In the problems we consider, the number of outputs is much larger
than the memory of the players. Therefore we use the following
output convention. The player who computes the value of an output
sends this value to the other player at a predetermined point in
the protocol, who is then allowed to forget the output. In order
to make the model as general as possible, we allow the players to
do local measurements, and to throw qubits away as well as pick up
some fresh qubits. The space requirement only demands that at any
given time no more than $S$ qubits are in use in the whole
circuit.

For more quantum background we refer to \cite{nc:qc}.

\subsection{The Discrepancy Lower Bound and Other Rectangle Bounds}\label{sec:rect}

\begin{definition}\label{def:ccm}
The communication matrix $M_f$ a function $f:\dom\to Z$ with rows and columns
corresponding to $X,Y$ is defined by $M_f(x,y)=f(x,y)$.

A rectangle is a product set in $\dom$. Rectangles are usually
labelled, an $\ell$-rectangle being labelled with $\ell\in Z$.
$\ell(R)$ gives the label of $R$.
\end{definition}

We will make use of the following simple observation.

\begin{proposition}\label{lem:triv}
Let $R\subseteq X^l\times Y^r$ be a rectangle. Then the set
\[R'[u,v]=\{x_i,y_j\,|\, u_1,\ldots,u_{i-1},x_i,u_{i+1},\ldots,u_l,
v_1,\ldots,v_{j-1},y_j,v_{j+1},\ldots,v_r\in R\}\] is a rectangle in $\dom$ for all fixed
values $u_a\in X$ and $v_b\in Y$, $1\le a,b\le l,r$.\end{proposition}

The discrepancy bound has been defined above. The application of
the discrepancy bound to communication complexity is as follows
(see \cite{kremer:thesis}):

\begin{fact}
A quantum protocol which computes a function $f:\fdom$ correctly
with probability $1/2+\epsilon$ over a distribution $\nu$ on the
inputs (and over its measurements) needs at least
$\Omega(\log(\epsilon/disc_\nu(f)))$ communication.
\end{fact}

We will use the following generalization of discrepancy to
matrices whose entries have more than two different values.

\begin{definition}
For a matrix $M$ with $M(x,y)\in Z$ for some finite set $Z$ we define its multicolor
discrepancy as \[mdisc(M)=\max_R \max_{z\in Z} |(\mu(R\cap f^{-1}(z))-\mu(R)/|Z|)|,\]
where the maximization is over all rectangles $R$ in $M$.
\end{definition}
The above definition corresponds to the notion of {\it strong}
multicolor discrepancy used previously in communication complexity
theory by Babai et al.~\cite{babaihk:help}. A matrix with high
multicolor discrepancy has rectangles whose measure of one color
is very different from the average $\mu(R)/|Z|$. Note that we have
defined this only for the uniform distribution $\mu$ here, and
that only functions for which all outputs have almost equal
probabilities are good candidates for small multicolor discrepancy
(e.g.~the inner product over finite fields).

We next define the one-sided rectangle bound on randomized
communication complexity, see Example~3.22 in
\cite{kushilevitz&nisan:cc} and also \cite{klauck:rect}.

\begin{definition}
Let $\nu$ be a distribution on $\dom$. Then $\nu$ is (strictly) balanced for $f:\fdom$,
if $\nu(f^{-1}(1))=1/2=\nu(f^{-1}(0)).$
\end{definition}

\begin{definition}
Let $err(R,\nu,\ell)=\nu(f^{-1}(1-\ell)|R)$ denote the error of an $\ell$-rectangle $R$.
Then let $size(\nu,\epsilon,f,\ell)=\max\{\nu(R)\,:\,err(R,\nu,\ell)\le \epsilon\}, $
where $R$ runs over all rectangles in $M_f$.

Define $bound^{(1)}_\epsilon(f)=\max_\nu\log(1/size(\nu,\epsilon,f,1)),$ where $\nu$ runs
over all balanced distributions on $\dom$. Finally,
\[bound(f)=\max\{bound^{(1)}_{1/4}(f),bound^{(1)}_{1/4}(\neg f)\}.\]
\end{definition}
The application to classical communication is as follows.

\begin{fact}
For any function $f:\fdom$, its (public coin) randomized
communication complexity with error 1/4 is lower bounded by
$bound(f)$.
\end{fact}

\section{Proving Quantum Communication-Space Tradeoffs}\label{sec-howto}

Suppose we are given a communicating quantum circuit that computes
$f_{l,r}$, i.e., the Alice circuit gets $l$ inputs from $X$, the
Bob circuit gets $r$ inputs from $Y$, and they compute all outputs
$f(x_i,y_j)$. Furthermore we assume that the output for pair
$(i,j)$ is produced at a fixed gate in the circuit.

Our approach to prove the lower bound is by slicing the circuit. Let $mdisc(f)=1/2^d$.
Then we partition the circuit in the following way. The first slice starts at the
beginning, and ends when $d/100$ qubits have been communicated, i.e., after $d/100$ qubit
wires have crossed between the Alice and Bob circuits. The next slice starts afterwards
and also contains $d/100$ qubits communication and so forth. Note that there are $O(C/d)$
slices, and $lr$ outputs, so an average slice has to make about $lrd/C$ outputs. We will
show that every such slice can produce only $O(S)$ output bits. This implies the desired
lower bound.

So we consider what happens at a slice. A slice starts in some state on $S$ qubits that
has been computed by the previous part of the computation. Then the two circuits run a
protocol with $d/100$ qubits communication. We have to show that there can be at most
$O(S)$ output bits. At this point the following observation will be helpful.

\begin{proposition}\label{lem:union}
Suppose there is an algorithm that on input $x$ first receives $S$
qubits of initial information depending arbitrarily on $x$ for
free. Suppose the algorithm produces some output correctly with
probability $p$.

Then the same algorithm with the initial information replaced by
the totally mixed state has success probability at least $p/2^S$.
\end{proposition}

Suppose the circuit computes the correct output with probability 1/2. Then each circuit
slice computes its outputs correctly with probability 1/2. Proposition~\ref{lem:union}
tells us that we may replace the initial state on $S$ qubits by a totally mixed state,
and still compute correctly with probability $(1/2)\cdot 1/2^S$. Hence it suffices to
show that any protocol with communication $d/100$ that attempts to make $\ell$ bits of
output has success probability exponentially small in $\ell$. Then $\ell$ must be bounded
by $O(S)$. What is left to do is provided by the following bipartite product result.

\begin{theorem}\label{the:prod1}
Suppose a quantum protocol with communication $d/100$ makes $k\le
d/(100\log|Z|)$ outputs for function values $f(x_i,y_j)$ of
$f:\dom\to Z$ with $mdisc(f)\le 2^{-d}$. Then the probability that
these outputs are simultaneously correct is at most
$(1+o(1))\cdot|Z|^{-k}$.
\end{theorem}

We establish this result in two steps. First we show that for each
function with multiple outputs and small multicolor discrepancy all quantum
protocols have small success probability.

\begin{lemma}\label{lem:protmdisc}
If there is a quantum protocol with communication $c$ that computes the outputs of a
function $f:X^l\times Y^r\to Z^k$ so that the success probability of the protocol is
$1/|Z|^k+\alpha$ (in the worst case), then $mdisc(f)\ge\Omega(\alpha^5/2^{10c})$.
\end{lemma}

Conversely, if $c\le-\log mdisc(f)/10-k\log |Z|$, then the success probability of quantum
protocols with communication $c$ is at most $(1+o(1))\cdot|Z|^{-k}$.

The next step is to derive multicolor discrepancy bounds for
$f_{l,r}$ from multicolor discrepancy bounds for $f$.

\begin{lemma}\label{lem:discmdisc}
Let $f:\dom\to Z$ have $mdisc(f)\le 2^{-d}$. Let the set
$O=\{(i_1,j_1),\ldots,(i_k,j_k)\}$ contain the indices of $k$
outputs for $f_{l,r}$. Denote by $f_O$ the function that computes
these outputs. Then $mdisc(f_O)\le O(2^{-d/4})$, if $k\le
d/5$.
\end{lemma}

These two lemmas imply Theorem~\ref{the:prod1}. Their proofs are in Appendix~\ref{app:A}
resp.~Appendix~\ref{app:B}. Now we can conclude the following more general version of
Theorem~\ref{the:cst}.

\begin{theorem}\label{the:main}
Let $f:\dom\to Z$ with $mdisc(f)\le 1/2^d$. Then every quantum
protocol using space $S$ that computes $f_{l,r}$ needs
communication $\Omega(dlr\log|Z|/S)$.
\end{theorem}

\begin{proof}
Note that if $S=\Omega(d)$, we are immediately done, since communicating the outputs
requires at least $lr\log|Z|$ bits. If $S\le d/200$, we can apply Theorem~\ref{the:prod1}
and Proposition~\ref{lem:union}. Consider a circuit slice with communication $d/100$ and
$\ell$ outputs. Apply Theorem~\ref{the:prod1} to obtain that the success probability of
any protocol without initial information is at most $(1+o(1))\cdot|Z|^{-k}$ for $k$ being
the minimum of $\ell$ and $d/(100\log|Z|)$. With Proposition~\ref{lem:union} we get that
this must be at least $(1/2)\cdot 2^{-S}$, and hence $k\le (S+2)/\log |Z|$. In the case
$k=d/(100\log|Z|)$ we get the contradiction $S+2\ge k\log |Z|=d/100$ to our assumption,
otherwise we get $\ell\le (S+2)/\log|Z|$ and hence $C/(d/100)\cdot (S+2)/\log|Z|\ge lr$
as desired.
\end{proof}

We also get the following corollary in the same way.

\begin{corollary}\label{cor:mcs}
Let $f$ be a function with $m$ output bits so that for all $k<d$
and each subset $O$ of $k$ output bits $mdisc(f_O)<2^{-d}$. Then every
quantum protocol with communication $C$ and space $S$ satisfies
the tradeoff $CS=\Omega(dm)$.
\end{corollary}

\section{Applications}

In this section we apply Theorem~\ref{the:main} and Corollary~\ref{cor:mcs} to show some
explicit communication-space tradeoffs. We have already stated our result regarding
matrix and matrix-vector products over finite fields in the introduction
(Corollary~\ref{cor:mmfield}). The only missing piece is an upper bound on the multicolor
discrepancy of $IP^F$ for finite fields $F$.

\begin{lemma}
$mdisc(\widetilde{IP}^F)\le |F|^{-n/4}$.
\end{lemma}

\begin{proof}
The following is proved in \cite{mnt:hash}.
\begin{fact}\label{fac:hash}
Let $Y$ be a pairwise universal family of hash functions from $X$
to $Z$. Let $A\subseteq X$, $B\subseteq Y$, and $E\subseteq Z$.
Then\begin{equation}\label{eq:mnt} \left|Prob_{x\in A,h\in
B}(h(x)\in E)-\frac{|E|}{|Z|}\right|\le
\sqrt{\frac{|Y|\cdot|E|}{|A|\cdot|B|\cdot|Z|}}.\end{equation}
\end{fact}

$IP^F$ can be changed slightly to give a universal family, with
$X=F^n$ and $Z=F$, by letting $h(x)=IP^F(x,y)+a$ for $y$ drawn
randomly from $F^n$ and $a$ from $F$. Then the set of hash
functions has size $|Y|=|F|^{n+1}$.

To bound the multicolor discrepancy of evaluating the hash family we can set $E$ to
contain any single element of $F$. Hence for each rectangle $A\times B$ containing at
least $|F|^{(3/2)\cdot n}$ entries the right hand side of inequality~(\ref{eq:mnt}) is at
most $|F|^{(n+1)/2}/(\sqrt{|F|^{(3/2) \cdot n}\cdot |F|})= |F|^{-n/4}$. This is an upper
bound on $\mu(A\times B)$ times the multicolor discrepancy, and hence also an upper bound
on the latter itself. Smaller rectangles can have multicolor discrepancy at most
$|F|^{-n/2-1}$, thus the multicolor discrepancy of evaluating the hash function is at
most $|F|^{-n/4}$. Hence also $IP^F$ has small discrepancy: its communication matrix is a
rectangle in the communication matrix for the hash evaluation.
\end{proof}

\begin{proof}[Proof of Corollary~\ref{cor:hash}.]
We again make use of Fact~\ref{fac:hash}. Assume that the output is encoded in binary in
some standard way using $\lceil\log |Z|\rceil$ bits. Fix an arbitrary value of $k$ output
bits to get a subset $E$ of possible outputs in $Z$. We would like to have
$|E|/|Z|=2^{-k}$, but this is not quite possible, e.g.~for $Z$ being $\{0,\ldots,p-1\}$
for some prime $p$. If we restrict ourselves to the lower $\log(|Z|)/2$ bits of the
binary encoding of elements of $Z$, however, then each such bit is $1$ resp.~0 with
probability $1/2\pm1/\sqrt{|Z|}$ for a uniformly random $z\in Z$, even conditioned on
other bits, so that the probability of a fixed value of $k$ of them is between
$(1/2-1/\sqrt{|Z|})^k$ and $(1/2+1/\sqrt{|Z|})^k$. Then $|\,\,|E|/|Z|-1/2^k\,\,|\le
2/\sqrt{|Z|}$.

Let $R=A\times B$ be any rectangle in the communication matrix. Assume that $|R|\ge
\sqrt{|X|}\cdot|Y|$. Then the right hand side of~(\ref{eq:mnt}) is
$\le\sqrt{|Y|/(\sqrt{|X|}|Y|)}=1/|X|^{1/4}$. If $R$ is smaller, then its multicolor
discrepancy is at most $1/\sqrt{|X|}$. So we can apply Corollary~\ref{cor:mcs} with a
multicolor discrepancy of at most $|X|^{-1/4}+2|Z|^{-1/2}$. Note that the number of
output bits we consider is $\log|Z|/2$, and we get $CS=\Omega(\log|X|\cdot\log|Z|)$ or
$\Omega((\log|Z|)^2)$, whichever is smaller.
\end{proof}

\section{A Direct Product Result for the Rectangle Bound}

Theorem~\ref{the:disjdpt} is an immediate consequence of the following direct product
result for the rectangle bound, plus a result of Razborov \cite{razborov:disj}.

\begin{lemma}
Let $f:\fdom$ be a function and denote by $f_k$ the problem to
compute $f$ on $k$ distinct instances. Assume that $bound(f)\ge b$ and
that this is achieved on a balanced distribution $\nu$.

Then there is a constant $\gamma>0$ such that the average success
probability of each classical protocol with communication $b/3$
for $f_k$ on $\nu^k$ is at most $2^{-\gamma k}$ for any $k\le b$.
\end{lemma}

The lemma is proved in Appendix~\ref{app:C}. Now we state the result of
Razborov~\cite{razborov:disj}.

\begin{fact} $bound(DISJ)\ge \epsilon n$ for some constant $\epsilon>0$.\end{fact}

Note that the distribution used in \cite{razborov:disj} is not strictly balanced, but can
be changed to such a distribution easily.

\bibliographystyle{alpha}

\newcommand{\etalchar}[1]{$^{#1}$}

\begin{appendix}

\section{Efficient Quantum Protocols and Small Multicolor Discrepancy}\label{app:A}

In this section we show Lemma~\ref{lem:protmdisc}. We first need the following fact from
\cite{klauck:qclb}. This result is proved by first decomposing a quantum protocol for
each of the possible values of all outputs into few rank one matrices, whose sum
expresses the probability of this particular output on the inputs in the communication
matrix. Then these matrices are discretized into rectangles (similar to
\cite{yao:qcircuit,kremer:thesis}). Note that we can assume that the protocol always has
a pure global state here, since we have dropped any space restrictions. Also the players
do not share any entanglement at the beginning of the protocol, because this is destroyed
by replacing the initial state of circuit slices by the totally mixed state on $S$ qubits
(which in turn can be replaced by $S$ qubits from a pure state on $2S$ qubits).

\begin{fact}
Assume there is a quantum protocol with communication $c$ that
computes a value of the output $O$ from a finite set $Z$ on each
input $x,y$ with probability $p_{x,y}$.

Then for each $\beta\ge 0$ there is a real $w\in[0,1]$, and a set of $O(2^{10c}/\beta^4)$
rectangles $R(i)$ with weights $w(R(i))\in\{-w,w\}$, so that
\[\sum_{i:(x,y)\in R(i)} w(R(i))\in [p_{x,y}- \beta,p_{x,y} ].\]
\end{fact}

Hence for our protocol that computes the $k$ outputs of $f$ we can
find $|Z|^k$ sets $M(b_1,\ldots, b_k)$ of rectangles, so that for
an input $x,y$ summing the weights of those rectangles in
$M(b_1,\ldots, b_k)$ that contain $x,y$ gives approximately the
probability of output $(b_1,\ldots, b_k)$ occurring when the
protocol runs on $x,y$. For $R\in M(b_1,\ldots, b_k)$ we define
$\ell(R)=(b_1,\ldots, b_k)$ as its label. Let $M$ be the union of
all the $M(b_1,\ldots,b_k)$.

Assume the protocol has an advantage of $\alpha$ over a random
guess for every input. Set $\beta=\alpha/2$ when applying the above
lemma. For every output value $b_1,\ldots, b_k$ and for every
input $x,y$ such that $f(x,y)=(b_1,\ldots, b_k)$ we have
\[\sum_{R\in M: \ell(R)=(b_1,\ldots,b_k), (x,y)\in R} w(R)
-1/|Z|^k\ge\alpha-\beta=\alpha/2,\] and thus, defining
$\delta(x,y,b_1,\ldots, b_k)=1-1/|Z|^k$ if $f(x,y)=(b_1,\ldots,
b_k)$ and $-1/|Z|^k$ otherwise,
\[\sum_{R\in M: (x,y)\in R} w(R) \delta(x,y,\ell(R)) \ge\alpha/2 ,\]
since for all $x,y: \sum_{R\in M: (x,y)\in R} w(R)\le 1$.

Hence, by averaging,
\[\sum_{x,y} \mu(x,y)\sum_{R\in M: (x,y)\in R} w(R) \delta(x,y,\ell(R)) \ge\alpha/2,\]
and by exchanging sums \[ \sum_{R\in M} w(R)\left(\mu(R\cap
f^{-1}(\ell(R)))-\mu(R)/|Z|^k\right)\ge\alpha/2.\] Consequently
there must be a rectangle $R$ with
\[|\mu(R \cap f^{-1}(\ell(R)))- \mu(R)/|Z|^k|\ge\frac{\alpha}{2}/(O(2^{10c}/\beta^4)=\Omega(\alpha^5/2^{10c}).\]
This rectangle has the stated multicolor discrepancy.

\section{A Bipartite Product Result for Multicolor Discrepancy}\label{app:B}

In this section we prove Lemma~\ref{lem:discmdisc}. Suppose that
$mdisc(f)\le 1/2^d$. Let $O=\{(i_1,j_1),\ldots,(i_k,j_k)\}$ be the
set of output labels, i.e., output $(i,j)\in O$ should correspond
to $f(x_i,y_j)$. Fix some function values $c_{i,j}$ for the
elements of $O$, i.e., $f(x_i,y_j)=c_{i,j}$. We want to show that
each rectangle in $\lrdom$ is either very small or contains a
fraction of inputs satisfying these constraints that is not much
larger than $1/|Z|^k$.

Fix some rectangle $R\subseteq\lrdom$. The probability when picking
$x,y=x_1,\ldots,x_l,y_1,\ldots y_r\in R$ that $f(x_i,y_j)=c_{i,j}$
for all $(i,j)\in O$ can be written as the product of conditional
probabilities
\begin{equation}\label{eq:prod}
\prod_{(i,j)\in O} Prob_{x,y\in
R}(f(x_i,y_j)=c_{i,j}\,|\,f(x_a,y_b)=c_{a,b}\mbox{ for }
(a,b)<(i,j); (a,b)\in O),\end{equation} assuming some order $<$ on
pairs $(i,j)$. For any single term in this product there are three
types of conditions: $(a,b)$ may satisfy
\begin{enumerate}
\item $a\neq i;b\neq j$, \item $a=i;b\neq j$,
 \item $a\neq i ; b=j$.
\end{enumerate}

The first type of condition involves neither $x_i$ nor $y_j$, the
others involve exactly one of them. We can write a term of
(\ref{eq:prod}) as
\begin{equation}\label{eq:condexp}
E_{u,v}Prob_{x,y\in R}(f(x_i,y_j)=c_{i,j}\,|\,x_a=u_a,y_b=v_b\mbox{ for } (a,b)\neq
(i,j),\mbox{ and } {\cal C}),
\end{equation}
where the distribution on $u,v\in X^{l-1}\times Y^{r-1}$ is given by picking $u_1,\ldots,u_l$, $v_1,\ldots v_r$
uniformly from those inputs in $R$ that satisfy $f(u_a,v_b)=c_{a,b}$ for all $(a,b)<(i,j);
(a,b)\in O$, and then dropping $u_i,v_j$.
${\cal C}$ denotes the conditions of the second and third type.

For all $a,b$ with $a\neq i,b\neq j$ fix {\it any} value of $x_a=u_a$ and $y_b=v_b$,  so
that $f(x_a,y_b)=c_{a,b}$ if $(a,b)<(i,j)$ and $(a,b)\in O$  (i.e., consider a term in
expectation (\ref{eq:condexp})). Now we are only left with the conditions of the other
two types. Also we obtain a rectangle $R'[u,v]$ in $\dom$, since all inputs but $x_i,y_j$
are fixed (see Proposition~\ref{lem:triv}).

The second and third types of conditions involve either $x_i$ or
$y_j$. Observe that each such condition
$f(x_i,y_b)=f(x_i,v_b)=c_{i,b}$ partitions $X$ into those $x_i$
satisfying it and those who do not, and
$f(x_a,y_j)=f(u_a,y_j)=c_{a,j}$ partitions $Y$. There are at most
$k$ such conditions, so all $m$ possible truth values of these
conditions partition $\dom$ into disjoint rectangles
$M_1,\ldots,M_m$ with $m\le 2^k$. We are interested in the measure
of inputs with $f(x_i,y_j)=c_{i,j}$ on those $x_i,x_j$ satisfying
the given conditions, i.e., lying in one of these $2^k$
rectangles. Hence we need to bound
\[Prob_{x,y\in R'[u,v]}(f(x,y)=c_{i,j}\,|\,x,y\in M_\ell)\]
for one of these rectangles $M_\ell$.

Suppose that $\mu(R'[u,v]\cap M_\ell)\le 2^{-d/2}$. All such
rectangles $R'[u,v]\cap M_\ell$ together contribute at most
$k\cdot 2^k \cdot 2^{-d/2}$ to the multicolor discrepancy of $R$,
as we argue now. The size of $R$ can be written as follows ($\mu$
is the uniform distribution on implicit domains).

\begin{eqnarray*}
\mu(R)&=&\sum_{u,v\in R}\mu(u,v)\\
&=&\sum_{u_1,\ldots,u_{l},v_{1},\ldots,v_{r}\in R}\left(\prod_{a\neq i, b\neq j} \mu(u_a)\mu(v_b)\right)
\mu(u_i)\mu(v_j)\\
&=&E_{u,v} \mu(R'[u,v])\\
&=& \sum_{1\le\ell\le m}  E_{u,v}\mu(R'[u,v]\cap M_\ell),
\end{eqnarray*}
where the expectation $E_{u,v}$ is over the distribution in which $u_1,\ldots,u_l,v_1,\ldots, v_r$ are
picked uniformly from $R$, and then $u_i$ and $v_j$ are dropped.

Hence if we ignore all the small rectangles $R'[u,v]\cap M_\ell$
in $R$ the measure of ignored inputs is at most $km2^{-d/2}$: for
each of the $k$ terms in the expression~(\ref{eq:prod}) and the $m$
possible outputs the above expectation cannot gain more than measure $2^{-d/2}$ from these rectangles.

So assume that $\mu(R'[u,v]\cap M_\ell)\ge 2^{-d/2}$ always.
$R'[u,v]\cap M_\ell$ is a rectangle in $\dom$, and hence has
multicolor discrepancy at most $2^{-d}$. Recall that we are interested in
\begin{eqnarray*}
&&Prob_{x,y\in R'[u,v]}(f(x,y)=c_{i,j}\,|\,x,y\in M_\ell)\\
&=& \mu(R'[u,v] \cap M_\ell\cap f^{-1}(c_{i,j}))\,/\, \mu(R'[u,v]\cap M_\ell),
\end{eqnarray*}
and so with \[\mu(R'[u,v] \cap M_\ell\cap f^{-1}(c_{i,j}))- \mu(R'[u,v]\cap M_\ell) /|Z|\le 2^{-d},\]
and $\mu(R'[u,v]\cap M_\ell)\ge 2^{-d/2}$ we get
\[Prob_{x,y\in R'[u,v]}(f(x,y)=c_{i,j}\,|\,x,y\in M_\ell)\le 1/|Z|+2^{-d/2}.\]

So, ignoring small rectangles that altogether contribute at most
$k2^k2^{-d/2}$ multicolor discrepancy, we have that each term in
the product of probabilities~(\ref{eq:prod}) is at most $1/|Z|+2^{-d/2}$, and
hence the product is at most $(1/|Z|+2^{-d/2})^k\le|Z|^{-k}+2\cdot
2^{-d/2}$, since $(1/|Z|+\gamma)^k\le 1/|Z|^k+2\gamma$ for all
$0\le\gamma\le 1/2$ and $|Z|\ge 2$. So the multicolor
discrepancy is at most $O(k2^k2^{-d/2})\le O(2^{-d/4})$.

\section{A Direct Product Result for the One-Sided Rectangle Bound}\label{app:C}

In this section we give the proof of Theorem~\ref{the:disjdpt}.
This theorem is an immediate consequence of our direct
product result for the rectangle bound (restated here), plus the aforementioned
result of Razborov \cite{razborov:disj}.

\begin{lemma}
Let $f:\fdom$ be a function and denote by $f_k$ the problem to
compute $f$ on $k$ distinct instances. Assume that $bound(f)\ge b$ and
that this is achieved on a balanced distribution $\nu$.

Then there is a constant $\gamma>0$ such that the average success
probability of each classical protocol with communication $b/3$
for $f_k$ on $\nu^k$ is at most $2^{-\gamma k}$ for any $k\le b$.
\end{lemma}

\begin{proof}
Every randomized classical protocol with some success probability
$p$ on a fixed distribution can be replaced by a deterministic
protocol with the same success probability using standard
techniques (since the success probability of a randomized protocol
is an expectation over deterministic protocols). So assume we are
given a deterministic protocol with communication $c$ and success
probability $p$ for $f_k$. Such a protocol naturally leads to a
partition of $\kdom$ into $2^c$ rectangles labelled with the
common output of the protocol on the inputs in these rectangles.
In our case there are $2^{b/3}$ rectangles.

Since $\nu$ is (strictly) balanced, on $\nu^k$ each sequence of
$k$ function values is equally likely. We know that each rectangle
on $\dom$ that has size at least $1/2^b$ cannot contain more than
a fraction of $3/4$ of 1-inputs to $f$. Intuitively for each
sequence of $k$ outputs that has $\Omega(k)$ ones in it, the
correctness probability goes down by a factor of $3/4$ with each
1-output and is hence exponentially small in $k$.

Assume the following claim:

\begin{claim}
Each rectangle of size $\ge 2^{-b/2}$ in $\kdom$ can contain at most a
fraction of $2^{-\Omega(k)}$ of inputs having $f_k(x)=c$ for
every $c\in \{0,1\}^k$ with $|c|\in \{k/3,\ldots, k\}$.
\end{claim}

Due to a simple application of the Chernoff bound all but a
fraction of $1/2^{\Omega(k)}$ of the inputs have function values $c$ with
$|c|\ge k/3$. Then the overall correctness probability of the
protocol on $\nu^k$ is bounded from above by $1/2^{\Omega(k)}$,
since apart from the inputs with less than $k/3$ ones in the
function value all other inputs lie in rectangles that either have
an error of $1-1/2^{\Omega(k)}$ or are smaller than $2^{-b/2}$.
The latter rectangles have a combined measure of at most
$2^{b/3}\cdot 2^{-b/2}\le 2^{-\Omega(k)}$, given that $k\le b$.

Let us prove the claim. Consider a rectangle $R\subseteq\kdom$ of
size $2^{-b/2}$. Fix any output string $c\in\{0,1\}^k$ with at
least $k/3$ ones. We are interested in the measure of inputs on
$R$ that have $c$ as function value. Again we may write the
measure as a product of conditional probabilities
\begin{equation}\label{eq:prod2}
\prod_iProb(f(x_i,y_i)=c_i|f(x_j,y_j)=c_j\mbox{ for } j<i).\end{equation}
We skip all terms concerning the probability that $f(x_i,y_i)=0$.
So we are interested in the probability that $f(x_i,y_i)=1$
conditioned on $f(x_j,y_j)=c_j$ for all
$j<i$.

Each term may be written as follows.
\begin{equation}\label{eq:condexp2}
E_{u,v}Prob_{x,y\in R}(f(x_i,y_i)=1\,|\,x_j=u_j,y_j=v_j
\mbox{ for } j\neq i),\end{equation}
where the distribution on $u,v\in X^{k-1}\times Y^{k-1}$  is given
by picking $u_1,\ldots,u_k$, $v_1,\ldots,v_k$
uniformly from those inputs in $R$ that satisfy
$f(u_j,v_j)=c_{j}$ for all $j<i$, and then dropping $u_i,v_i$.

Again fix all $x_j,y_j$, $j\neq i$ in any way under the condition
that $f(x_j,y_j)=c_j$ for all $j<i$. This leaves us
with a rectangle $R'[u,v]\subseteq\dom$.

If $\nu(R'[u,v])<2^{-b}$, then we ignore $R'[u,v]$, since all such $R'[u,v]$ together will
not influence the error of $R$ significantly. More precisely,
since all rectangles $R'[u,v]$ obtained by fixing $u,v$ are
disjoint parts of $R$ when extended by $u,v$ to rectangles in $\kdom$,
the combined size of all these small rectangles
on $\kdom$ is at most $2^{-b}$. All rectangles ignored in this way
in any of the at most $k$ product terms in~(\ref{eq:prod2}) together have weight at
most $k2^{-b}$, which is at most a $k2^{-b}/2^{-b/2}$ contribution
relative to $R$.

So assume that $\nu(R'[u,v])\ge 2^{-b}$ always. Then clearly
$R'[u,v]$ contains at most a fraction of $3/4$ of 1-inputs to $f$,
by the definition of $bound(f)$.
All inputs with output $c$ need to have a 1 as output on block
$i$. Hence the fraction of inputs with output $c$ on $R$ is at
most $(3/4)^{k/3}+k\cdot2^{-b}/2^{-b/2}=2^{-\Omega(k)}$, since
$k\le b$.
\end{proof}

\section{Classical Communication-Space Tradeoffs}\label{app:D}

\begin{proof}[Proof of Theorem~\ref{the:disjtrade}.]
First we prove the lower bound for matrix multiplication. We may assume that $S\le
\gamma\sqrt{\epsilon n}/2$, for the constants from Theorem~\ref{the:disjdpt}, because the
$n^2$ outputs are included in the communication and so otherwise we immediately have
$CS^2=\Omega(n^3)$. Consider circuit slices of a communicating circuit for matrix
multiplication, each circuit slice containing communication $\gamma\epsilon n/(2S)$. Let
$\ell$ denote the number of outputs in any slice. If $\ell < 2S/\gamma$ we will be able
to get the lower bound easily. So assume there are more outputs, and choose any
$k=2S/\gamma$ of them. Then we will apply Theorem~\ref{the:disjdpt} (details follow
later) to show that $k$ outputs can be computed only with success probability $2^{-\gamma
k}$, and hence $(1/2)\cdot 1/2^S\le 2^{-\gamma k}$ with Proposition~\ref{lem:union}. This
leads to the contradiction that $S+1\ge 2S$, hence the slice makes only $2S/\gamma$
outputs, and so $C\cdot (\gamma\epsilon n)^{-1}\cdot 2S\cdot 2S/\gamma\ge n^2$.

Consider a classical protocol with $\gamma\epsilon n/(2S)=\epsilon
n/k$ bits of communication. We partition the universe
$\{1,\ldots,n\}$ of the Disjointness problems to be computed into
$k$ mutually disjoint subsets $U(i,j)$ of size $n/k$, each
associated to an output $(i,j)$, which in turn corresponds to a
row/column pair $A[i]$, $B[j]$ in the input matrices $A$ and $B$.
Assume that there are $a$ outputs $(i,j_1),\ldots, (i,j_a)$
involving $A[i]$. Each output is associated to a subset of the
universe $U(i,j_t)$, and we set $A[i]$ to zero on all positions
that are not in one of these subsets. Then we proceed analogously
with the columns of $B$.

If the protocol computes on these restricted inputs, it has to
solve $k$ instances of Disjointness of size $n/k$ each, since
$A[i]$ and $B[j]$ contain a single block of size $n/k$ in which
both are not set to 0 if and only if $(i,j)$ is one of the $k$
outputs. Hence Theorem~\ref{the:disjdpt} is applicable. Note that
$k=2S/\gamma\le\sqrt{\epsilon n}$ and hence $k\le \epsilon n/k$ as
required.

The proof for matrix-vector product is analogous.
\end{proof}

\end{appendix}
\end{document}